# Table-like magnetocaloric effect and enhanced refrigerant capacity in EuO$_{1-\delta}$ thin films

P. Lampen[1], R. Madhogaria[1], N.S. Bingham[1], M.H. Phan[1,*], P.M.S. Monteiro[2], N.-J. Steinke[2], A. Ionescu[2], C.H.W. Barnes[2], and H. Srikanth[1,*]

[1]Department of Physics, University of South Florida, Tampa, FL 33624, USA

[2]Cavendish Laboratory, Physics Department, University of Cambridge, Cambridge CB3 0HE, UK

An approach to adjusting the conduction band population for tuning the magnetic and magnetocaloric response of EuO$_{1-\delta}$ thin films through control of oxygen vacancies ($\delta = 0$, 0.025, and 0.09) is presented. The films each showed a paramagnetic to ferromagnetic transition around 65 K, with an additional magnetic ordering transition at higher temperatures in the oxygen deficient samples. All transitions are observed to be of second order. A maximum magnetic entropy change of 6.4 J/kg K over a field change of 2 T with a refrigerant capacity of 223 J/kg was found in the sample with $\delta = 0$, and in all cases the refrigerant capacities of the thin films under study were found to exceed that reported for bulk EuO. Adjusting the oxygen content was shown to produce table-like magnetocaloric effects, desirable for ideal Ericsson-cycle magnetic refrigeration. These films are thus excellent candidates for small-scale magnetic cooling technology in the liquid nitrogen temperature range.

## 1. Introduction

Magnetic refrigeration based on the magnetocaloric effect (MCE) has emerged as a promising alternative to the less efficient gas compression cycle methods.[1-4] However, the need for improved performance in magnetocaloric materials at moderate fields (up to 2 T, in the range of permanent magnets) and over a large temperature interval remains an obstacle to the wide-scale adoption of magnetic refrigeration technology.[3,4] A primary goal of research in the area of magnetic refrigeration is to identify materials that exhibit both large MCE (isothermal magnetic entropy change $\Delta S_M$ or adiabatic temperature change $\Delta T_{ad}$) and large refrigerant capacity (RC).[4] The RC is an important parameter to maximize for effective magnetic refrigeration devices as it provides a measure of the amount of heat transfer between the cold and hot sinks in an ideal refrigeration cycle. Its value depends on both the magnitude of $\Delta S_M$ and on its temperature dependence (i.e., the full width at half maximum of the $\Delta S_M(T)$ peak).

From a magnetic cooling perspective, exploring magnetocaloric materials in thin film form is of practical importance for small-scale cooling applications such as microelectromechanical systems (MEMS) and nanoelectromechanical systems (NEMS).[5-11] Relative to their bulk counterparts, thin films have the advantage of using less material, thus minimizing costs while the large surface to volume ratio reduces the time taken by the heat exchange process.[12,13] The Curie temperature's distribution in thin film systems, due to variation in the local magnetic environment around substrate-induced defects, may broaden the $\Delta S_M(T)$ curve over a wide temperature range, thus enhancing $RC$.[14-16] In many cases however, diminished magnetization emerges as a consequence of reduced dimensionality, which can lead to a significant drop in peak $\Delta S_M$. For example, the canonical room temperature MCE material Gd exhibits a peak value of $-\Delta S_M = 2.05$ J/kg K over an applied magnetic field change of 1 T;

this value was reduced by 20 – 60% in thin film form depending on preparation conditions,[17] and a study of Gd/W heterostructures showed a 47% drop in $-\Delta S_M$ as compared with bulk Gd.[18] The decline in peak entropy change is even more pronounced in manganite systems.[15,19] A precipitous drop in $\Delta S_M$ can negate the potential advantages of a broadened MCE curve and render such a film impractical for active magnetic refrigeration applications. Thus, it continues to remain challenging to enhance $RC$ while retaining large MCE in thin film systems in spite of their desirable properties.

EuO is a semiconducting ferromagnet with a $4f^7$ electronic configuration, yielding a large magnetic moment of 7 $\mu_B$ with a $T_C$ near 70 K.[20-24] The resistivity drops by up to 13 orders of magnitude when crossing into the ferromagnetic region, accompanied by nearly 100% spin polarization, a desirable property for spintronic applications.[20,25] An investigation of the MCE in bulk polycrystalline EuO by Ahn et al. showed a large peak value of magnetic entropy change of $-\Delta S_M$ = 8.4 J/kg K and 17.5 J/kg K for applied fields of 2.0 T and 5.0 T, respectively.[26] EuO was also utilized in composite materials with clathrates to produce a table-like magnetocaloric effect with very large refrigerant capacity.[27] Electron doping has been shown to enhance the Curie temperature up to 140 K as the excess electrons provided by the doping mediate an additional ferromagnetic exchange interaction between the $Eu^{2+}$ ions, an effect which has typically been achieved with the incorporation of other rare earth elements or trivalent transition metals.[20,25,28] Another approach is to adjust the conduction band population by controlling the oxygen vacancies in $EuO_{1-\delta}$, which act as dopants.[29,30]

In this paper, we demonstrate the possibility of tuning the magnetic and magnetocaloric responses of $EuO_{1-\delta}$ thin films by controlling the oxygen vacancy ($\delta$). A comparative study of the magnetic and magnetocaloric properties of the films with $\delta$ = 0, 0.025, and 0.09 has been performed. We have achieved the enhancement of RC by up to 30% in the films relative to their bulk counterpart, as well as a table-like MCE in the oxygen-deficient films.

## 2. Experimental

Thin films of EuO$_{1-\delta}$ with $\delta$ = 0, 0.025, and 0.09 were deposited by magnetron cosputtering of ceramic Eu$_2$O$_3$ and metallic Eu targets on Si substrates with a Pt buffer and capping layer. The polycrystalline samples were deposited at room temperature with a base pressure of 5×10$^{-9}$ Torr. Oxygen stoichiometry was controlled by varying the DC deposition current for the Eu target between 0.05 A and 0.15 A while maintaining the RF power at 75 W for the Eu$_2$O$_3$ target. The nominal structure of the films was Si(001)/SiO$_2$(1.4 nm)/Pt(5 nm)/EuO$_{1-\delta}$(100 nm)/Pt(5 nm). Growth parameters were controlled to obtain samples with 0%, 2.5%, and 9% oxygen deficiency ($\delta$ = 0, 0.025, and 0.09). Magnetic measurements were performed using a commercial Quantum Design Physical Properties Measurement System (PPMS) with a vibrating sample magnetometer.

## 3. Results and Discussion

X-ray diffraction scans performed on a Bruker D8 diffractometer show no impurity phases for the sample with $\delta$ = 0 (Fig. 1(a)) within the detection limit. In addition to the EuO pattern, peaks associated with Eu$_2$O$_3$ are observed for $\delta$ = 0.025, and both Eu$_2$O$_3$ and Eu impurity phases are present in the $\delta$ = 0.09 sample. The estimated total volume of impurity phases in the oxygen deficient samples is about 10% (a better refinement was hindered by the fact that the stoichiometric EuO film shows a preferential (textured) orientation along the (111) direction, while the oxygen deficient films show (100) textures). A determination of the thickness of the EuO$_{1-\delta}$ layer was made through X-ray reflectivity (XRR) measurements,[31] yielding 97.9 nm and 89.3 nm for the $\delta$ = 0 and $\delta$ = 0.025 samples, respectively. A large interface roughness due to more rapid deposition of Eu prevented the acquisition of useful XRR data for the $\delta = 0.09$ sample, and the nominal value of 100 nm was used in volume calculations. The error in the thickness of this film can be conservatively approximated as ±10%.

The temperature dependence of in-plane magnetization (Fig. 2) reveals a single ferromagnetic transition in the stoichiometric sample with $T_C \sim 65$ K, close to the expected bulk value of 69 K.[20] In the oxygen deficient films two magnetic ordering transitions are observed, with the onset of magnetic order occurring near 144 K and 142 K for $\delta = 0.025$ and $\delta = 0.09$, respectively, as estimated from $dM/dT(T)$ (Inset: Fig 2(a)). This so-called "double-dome" feature has been previously observed in several studies of Eu-rich EuO,[28,30,32] although the origin of this behavior remains disputed. Some authors have invoked a chemical phase separation scenario or magnetic polarons bound to oxygen vacancies.[30,33] Alternatively, since oxygen vacancies act as electron donors and thus increase the population of the conduction band, the RKKY-like indirect exchange proposed to describe the magnetic interactions in oxygen deficient EuO could lead to the observed elevated transition temperature in the oxygen deficient samples.[34] A recent study performed by using muon spin rotation on these samples showed that magnetic polarons and phase separation can be ruled out at these doping levels and that the RKKY-like interaction is the only model compatible with the observed increase in $T_C$.[31] Under the application of increasing magnetic fields, a broadening of the temperatures spanned by each transition is evident, so that in the $M(T)$ curves measured in a 2 T applied field (Fig. 2(b)) the two magnetic transitions observed at 0.1 T for the $\delta = 0.025$ and $\delta = 0.09$ films are no longer distinct. Closely spaced ferromagnetic transitions can lead to a table-like magnetocaloric effect with a wide working temperature range and large $RC$. Usually, such sequential transitions are realized in composite samples combining phases with multiple $T_C$'s.[27] While the changes in magnetization across the two transitions in the present $EuO_{1-\delta}$ films are not similar in magnitude, the oxygen deficient samples present an interesting example of multiple ferromagnetic transitions in a single material, which is expected to impact the *RC* favorably.

An examination of $H/M$ vs. $M^2$ constructs across the temperature region of interest in each sample revealed the absence of negative slopes for any isotherm, satisfying the Banerjee criterion for materials undergoing continuous magnetic transitions. This confirms the second order nature of both the high and low temperature transitions in the $EuO_{1-\delta}$ films, which is preferred in magnetic refrigeration cycles due to the thermal hysteresis that often accompanies first order magnetic transitions. While the proximity of the two transitions in the oxygen deficient samples prevented additional analysis, the critical properties of the single transition in the stoichiometric film were determined based on the Arrott-Noakes equation of state, which predicts a linear dependence of $M^{1/\beta}$ on $(H/M)^{1/\gamma}$, where $\beta$ and $\gamma$ are the critical exponents governing magnetization ($T < T_C$) and susceptibility ($T > T_C$), respectively.[34] Figure 3 (a) shows an approximately linear behavior in the Arrott-Noakes plot for the $\delta = 0$ film when the critical exponents of the 3D Heisenberg model for isotropic ferromagnets ($\beta = 0.365$, $\gamma = 1.386$) were used as trial values. The spontaneous magnetization $M_S(T)$ and initial inverse susceptibility $(H/M)_0(T)$ were extracted from the intercepts of a linear fitting of the isotherms in the high field region. These quantities obey the asymptotic relations $M_S(T) \propto |\varepsilon|^{\beta}$ and $(H/M)_0(T) \propto |\varepsilon|^{\gamma}$, where $\varepsilon = (T - T_C)/T_C$ is the reduced temperature, allowing the critical exponents to be determined from a power law fitting (Fig. 3(b)). Taking the error into consideration, the fitted value of $\beta = 0.407 \pm 0.02$ is reasonably close to the 3D Heisenberg-like critical exponents observed previously in bulk EuO ($\beta = 0.368 \pm 0.005$)[36] and a single crystalline thin EuO film ($\beta = 0.35 \pm 0.01$).[37]

The magnetic entropy change was calculated from a series of $M(H)$ isotherms through the application of the thermodynamic Maxwell relation,

$$\Delta S_M = \mu_0 \int_0^{H_{max}} \left(\frac{\partial M}{\partial T}\right)_H dH,$$

where $M$ is the magnetization, $H$ is the magnetic field, and $T$ is the temperature. The temperature dependence of $-\Delta S_M$ is plotted in Fig. 4 for various applied field changes of up to 2 T. The main peak in the entropy change is found near the lower temperature transition identified in the $M(T)$ curves: ~66 K for the $\delta = 0$ and $\delta = 0.025$ films and ~62 K for the $\delta = 0.09$ film. Above the main peak, additional features can also be seen in the oxygen deficient films in the vicinity of the higher temperature transition, broadening the temperature range over which $-\Delta S_M$ is nonzero. The peak magnitude of $-\Delta S_M$ ($-\Delta S_M^{max}$) reaches 6.4 J/kg K in the stoichiometric EuO film for an applied field change of 2 T, a significant fraction of the 8.4 J/kg K entropy change in bulk EuO in contrast to manganite systems in which reductions in $-\Delta S_M^{max}$ as large as ~80% have been observed for thin films as compared to their bulk counterparts.[15,19] In the oxygen deficient films $-\Delta S_M^{max}$ decreases considerably as a consequence of reduced magnetic moment and broader transitions (Fig. 2). The growth of $-\Delta S_M^{max}$ with field (according to $-\Delta S_M^{max} \propto H^n$) is shown in the inset of Fig. 4(b). The scaling exponent $n$ is related to critical exponents $\beta$ and $\gamma$ by $n = 1 + [(1 + \gamma/\beta)(1 - 1/\beta)]^{-1}$. The expected value of $n = 0.81$ for the $\delta = 0$ sample based on the critical exponents calculated above is confirmed by a linear fitting of $\ln S$ vs. $\ln H$. The $\delta = 0.025$ film shows a similar growth of $-\Delta S_M^{max}$ with magnetic field, however the dependence is close to linear in the $\delta = 0.09$ sample.

The refrigerant capacity is given by $RC = \int_{T_1}^{T_2} -\Delta S_M dT$, where $T_1$ and $T_2$ are taken as the temperatures of the FWHM of the $-\Delta S_M(T)$ peak. The area corresponding to the $RC$ is illustrated in the inset of Fig. 4(a) for the $\delta = 0$ film. A magnetic field change of 2 T produces an $RC$ of 223 J/kg in this sample whereas similar conditions in bulk EuO yielded $RC = 170$ J/kg and 168 J/kg K.[26,27] The origin of the enhancement can be attributed to variation in the local magnetic environment around substrate-induced defects that gives rise to a slight distribution of $T_C$'s.[15] The broadened transition compensates for the reduced peak value of $-\Delta S_M(T)$ in the calculation of the $RC$, which depends on both the height and FWHM of the $-\Delta S_M(T)$ peak. The 2 T $RC$ in the oxygen deficient thin films is also larger than that of the bulk EuO, although less than that of the stoichiometric film (Fig. 5). It is significant that large $RC$ values are achieved in the present samples in the moderate field range ($\leq 2$ T), within the capabilities of permanent magnets, as the added expense and space requirements that accompany superconducting magnets render their use impractical for many applications. It is also interesting to note that the table-like MCE (i.e., the relatively constant $-\Delta S_M$ with temperature) observed for the oxygen deficient film ($\delta = 0.09$) over a wide temperature range (10 K – 80 K) makes it desirable for Ericsson-cycle based magnetic refrigeration.[38-44] The magnetic transitions in EuO fall within a temperature range suitable to the liquefaction of cryogenic fluids rather than commercial room temperature refrigeration, however the large enhancement (> 30%) of the $RC$ in thin films of EuO suggests a potential avenue toward improving this magnetocaloric figure of merit in known refrigerant materials in their respective working temperature ranges.

## 4. Conclusion

In summary, the magnetocaloric properties of thin films of $EuO_{1-\delta}$ with $\delta = 0$, 0.025, and 0.09 prepared by magnetron cosputtering were examined. While the stoichiometric film was observed to show a single ferromagnetic transition, the oxygen deficient samples underwent sequential ferromagnetic transitions, broadening the potential working range of the material. The peak magnetic entropy change in the thin film samples was reduced when compared with that of bulk EuO, however enhanced *RC* was observed in each of the present samples. For a moderate applied field change of 2 T the *RC* of 223 J/kg in the film with $\delta = 0$ represents an increase of ~ 31% over the reported bulk values. These excellent properties make the present films attractive for active magnetic refrigeration in the liquid nitrogen temperature range.

**Acknowledgments**

Research at the University of South Florida was supported by the U.S. Department of Energy, Office of Basic Energy Sciences, Division of Materials Sciences and Engineering under Award No. DE-FG02-07ER46438. P.M.S.M. thanks the EPSRC, U.K., and Fundação para a Ciência e a Tecnologia (SFRH/BD/71756/2010), Portugal. We thank Dr. A. Chaturvedi for his assistance with some magnetic measurements.

**Figure Captions**

**Fig. 1** X-ray diffraction patterns for the (a) $\delta = 0$, (b) $\delta = 0.025$, and (c) $\delta = 0.09$ EuO$_{1-\delta}$ thin films deposited on Si substrates and capped with a Pt layer.

**Fig. 2** Temperature dependence of magnetization in the series of EuO$_{1-\delta}$ films under an applied magnetic field of (a) 0.1 T and (b) 2.0 T. Insets: First derivatives of magnetization with temperature.

**Fig. 3** (a) Arrott-Noakes plot of magnetization isotherms acquired between 38 K and 98 K for the stoichiometric thin film using 3D Heisenberg trial values of the critical exponents $\beta$ and $\gamma$. (b) Power law fit to the saturation magnetization and initial susceptibility obtained by extrapolating linear fittings of the isotherms in (a) in the high field region back to the axes.

**Fig. 4** Temperature dependence of magnetic entropy change over various applied field changes in the EuO$_{1-\delta}$ films with (a) $\delta = 0$, (b) $\delta = 0.025$, and (c) $\delta = 0.09$. Inset (a): Representation of the *RC*. Inset (b): Log-log plot of peak magnetic entropy change vs. applied field change.

**Fig. 5** Comparison of the refrigerant capacity of the present EuO$_{1-\delta}$ thin film samples in a 2T applied field with previous results for bulk polycrystalline EuO reported in the literature.

**Figure 1**

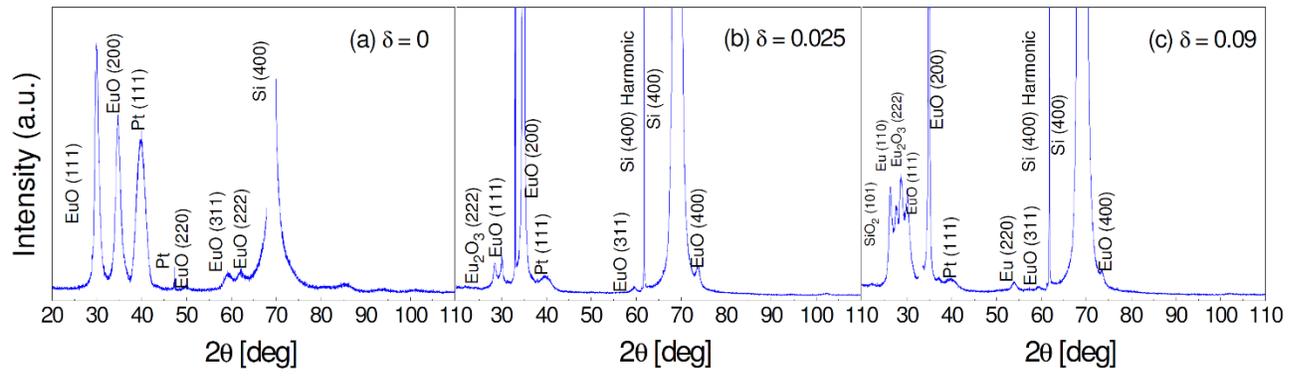

**Figure 2**

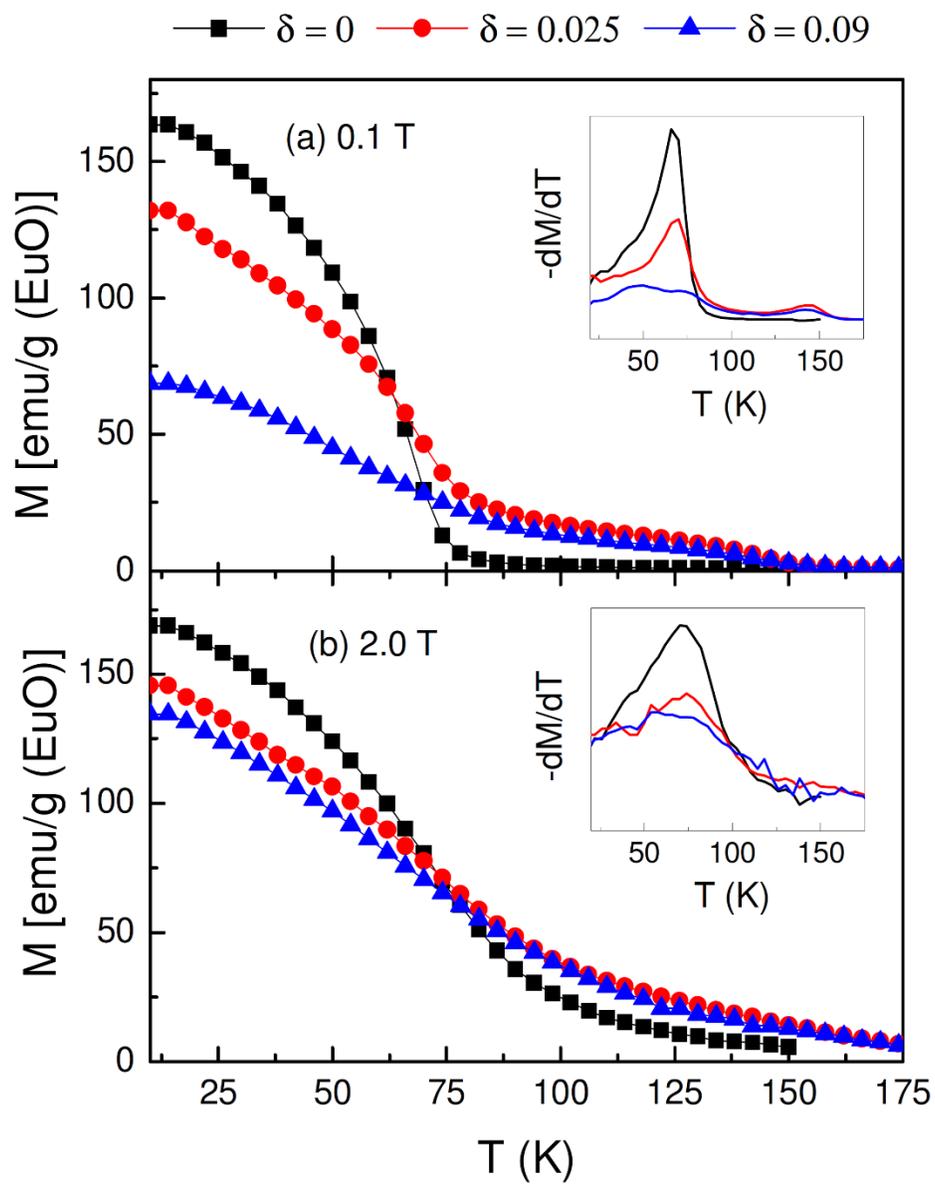

**Figure 3**

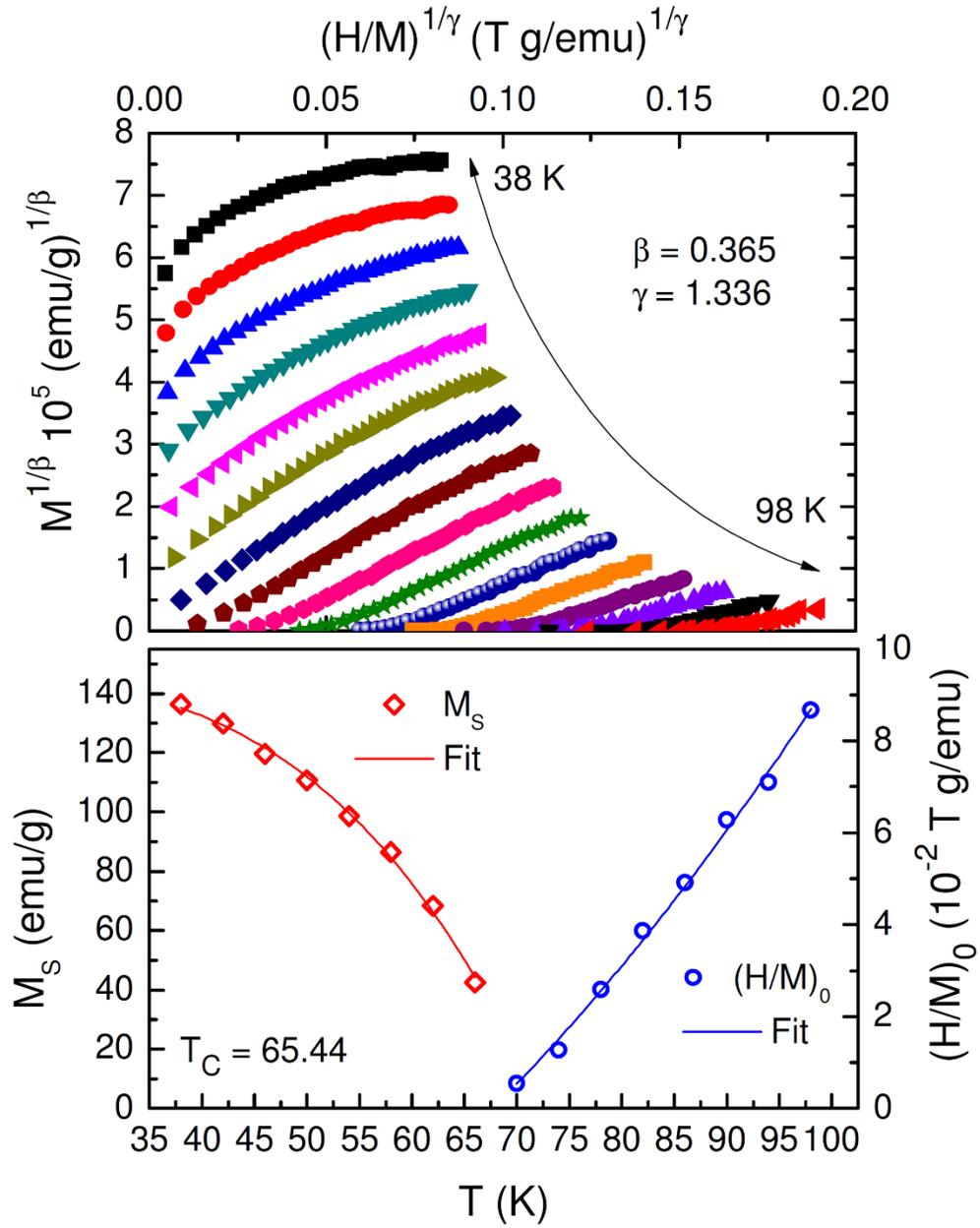

**Figure 4**

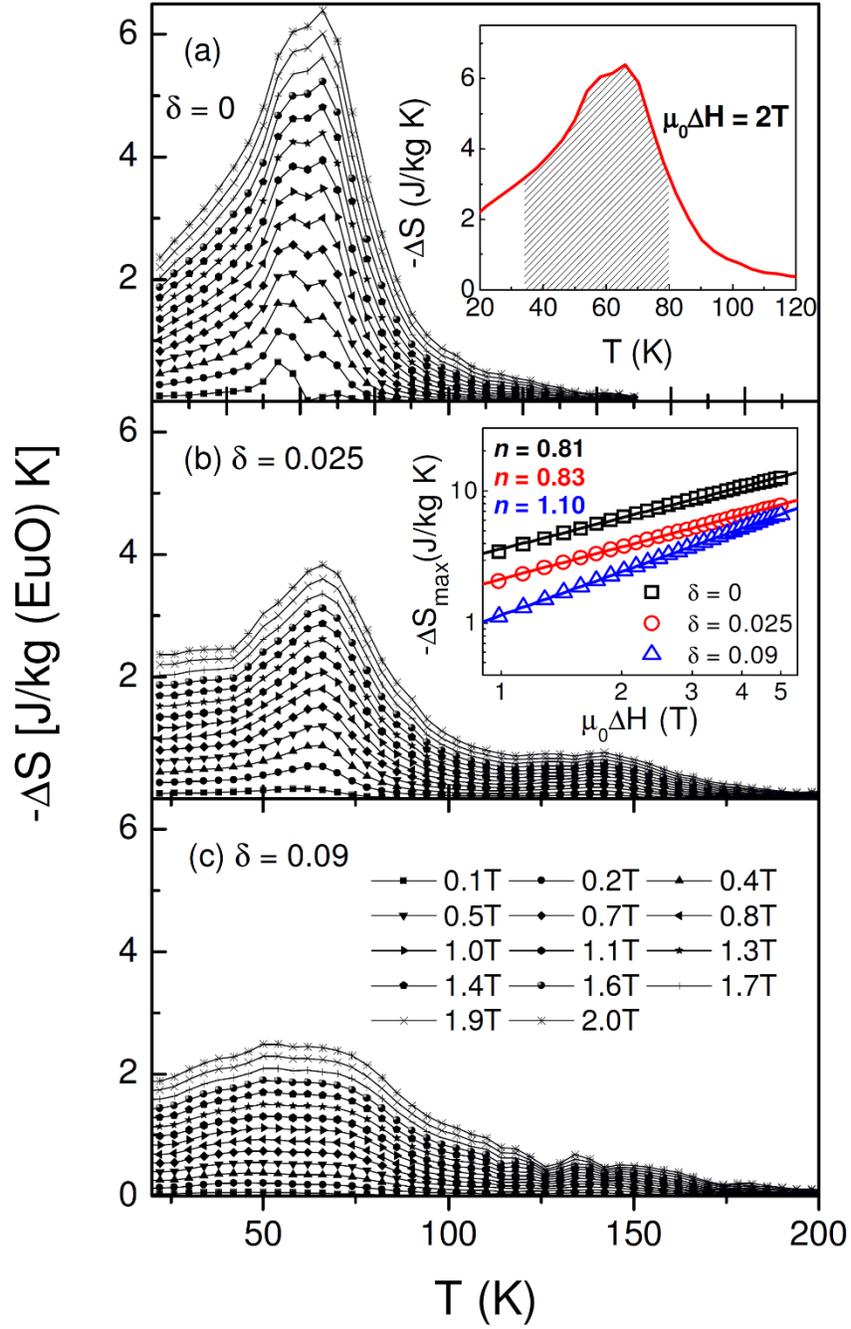

**Figure 5**

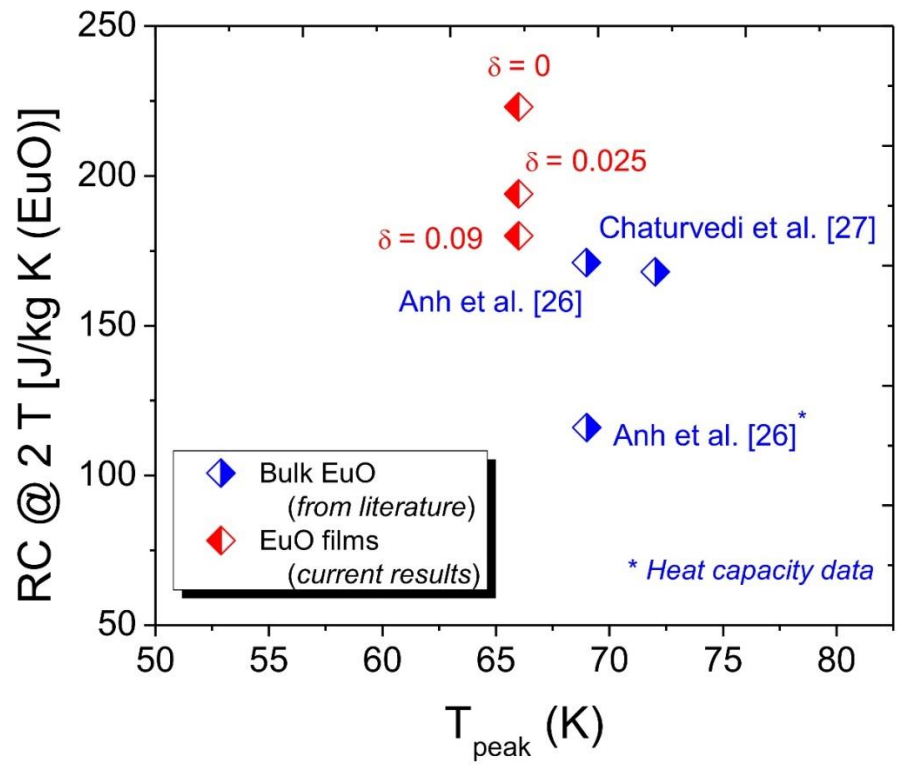